\documentclass[aps,prl,twocolumn,superscriptaddress,showpacs]{revtex4}
\usepackage{bbm}
\usepackage{txfonts}
\usepackage{graphicx}

\begin{document}

\title{Novel electronic structure induced by a highly strained oxide interface with incommensurate crystal fields}

\author{H. W. Ou}
\author{J. F. Zhao}
\author{Y. Zhang}
\author{B. P. Xie}
\author{D. W. Shen}
\affiliation{Department of Physics, Surface Physics Laboratory
(National Key Laboratory), Fudan University, Shanghai 200433, P. R.
China}

\affiliation{Advanced Materials Laboratory, Fudan University,
Shanghai 200433, P. R. China}

\author{Y. Zhu}
\author{Z. Q. Yang}
\author{J. G. Che}
\affiliation{Department of Physics, Surface Physics Laboratory
(National Key Laboratory), Fudan University, Shanghai 200433, P. R.
China}

\author{X. G. Luo}
\author{X. H. Chen}
\affiliation{Hefei National Laboratory for Physical Sciences at
Microscale and Department of Physics, University of Science and
Technology of China, Hefei, Anhui 230026, P. R. China}

\author{M. Arita}
\author{K. Shimada}
\author{H. Namatame}
\author{M. Taniguchi}
\affiliation{Hiroshima Synchrotron Radiation Center and Graduate
School of Science, Hiroshima University, Hiroshima 739-8526, Japan.}

\author{C. M. Cheng}
\author{K. D. Tsuei}
\affiliation{National Synchrotron Radiation Research Center, and
Department of Physics, National Tsing-Hua University, Hsinchu 30077,
Taiwan, Republic of China}

\author{D. L. Feng}
\email{dlfeng@fudan.edu.cn} \affiliation{Department of Physics,
Surface Physics Laboratory (National Key Laboratory), Fudan
University, Shanghai 200433, P. R. China}

\affiliation{Advanced Materials Laboratory, Fudan University,
Shanghai 200433, P. R. China}

\date{\today}

\begin{abstract}
The misfit oxide, Bi$_{2}$Ba$_{1.3}$K$_{0.6}$Co$_{2.1}$O$_{y}$, made
of alternating rocksalt-structured [BiO/BaO] layers and hexagonal
CoO$_{2}$ layers, was studied by angle-resolved photoemission
spectroscopy. Detailed electronic structure of such a highly
strained oxide interfaces is revealed for the first time. We found
that under the two incommensurate crystal fields, electrons are
confined within individual sides of the interface, and scattered by
umklapp scattering of the crystal field from the other side. In
addition, the high strain on the rocksalt layer raises its chemical
potential and induces large charge transfer to the CoO$_{2}$ layer.
Furthermore, a novel interface effects, the interfacial enhancement
of electron-phonon interactions, is discovered. Our findings of
these electronic properties lay a foundation for designing future
functional oxide interfaces.
\end{abstract}


\maketitle

Oxide interfaces have attracted much attention for the emergence of
novel phenomena
\cite{Ohtomo04,Nakagawa06,MgOZnO,Ueda98,LSMOYBCO,Sawatzky00,Reyren07}.
For example, high conductivity \cite{Ohtomo04} and even
superconductivity \cite{Reyren07} were discovered at the interface
between two insulators, LaAlO${_3}$ and SrTiO$_3$; ferromagnetism
was discovered in the superlattice of two antiferromagnets,
LaCrO$_3$ and LaFeO$_3$ \cite{Ueda98}; and gigantic thermoelectric
power at the SrTiO$_3$/SrTi$_{0.8}$Nb$_{0.2}$O$_3$ interface was
discovered five times higher than the bulk material\cite{Ohta}.
Clearly, it provides opportunities for promising applications: one
may be able to make oxide heterostructure devices with desired
physical properties\cite{Oxideinterface}. A thorough understanding
of the microscopic processes at the oxide interface is the first
step toward designing functional heterostructures. Particularly, it
is crucial to find out the electronic behavior at oxide interfaces,
\textit{e.g.}, how the electron bands react to two competing, and
sometimes incommensurate periodic potentials \cite{Voit} on both
sides of the interface; and how they react to strain, and peculiar
phonon structures at the interface \cite{MgOZnO,Locquet}. The
answers to these questions would lead to the understanding of the
charge transfer process, the insulator-metal transitions, and
anomalous transport behavior of the
interfaces\cite{Nakagawa06,LSMOYBCO,Millis}.

\begin{figure}[b]
\includegraphics[width=8cm]{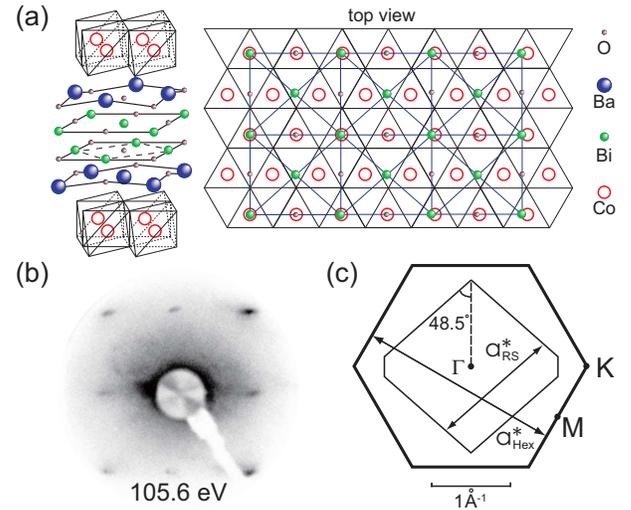}
\caption{(color online). (a) A schematic picture of the crystal
structure of Bi$_{2}$M$_{2}$Co$_{2}$O$_{y}$ (M= Ca, Sr, Ba). The
oxygen atoms in the CoO$_{2}$ layers, and BaO layer in the topview
are omitted for simplicity. (b) Low-energy electron diffraction
pattern exhibits the four-fold symmetry of the orthorhombic BiO
surface of Bi$_{2}$Ba$_{1.3}$K$_{0.6}$Co$_{2.1}$O$_{y}$. (c) The
reduced Brillouin zones (BZ) for the rocksalt [BiO/BaO] layers (thin
lines) and the hexagonal CoO$_{2}$ layer (thick lines). The
corresponding reciprocal lattice constants $a_{RS}^{*}$ and
$a_{Hex}^{*}$ are also illustrated.}
\end{figure}

\begin{figure*}[t!]
\includegraphics[width=15cm]{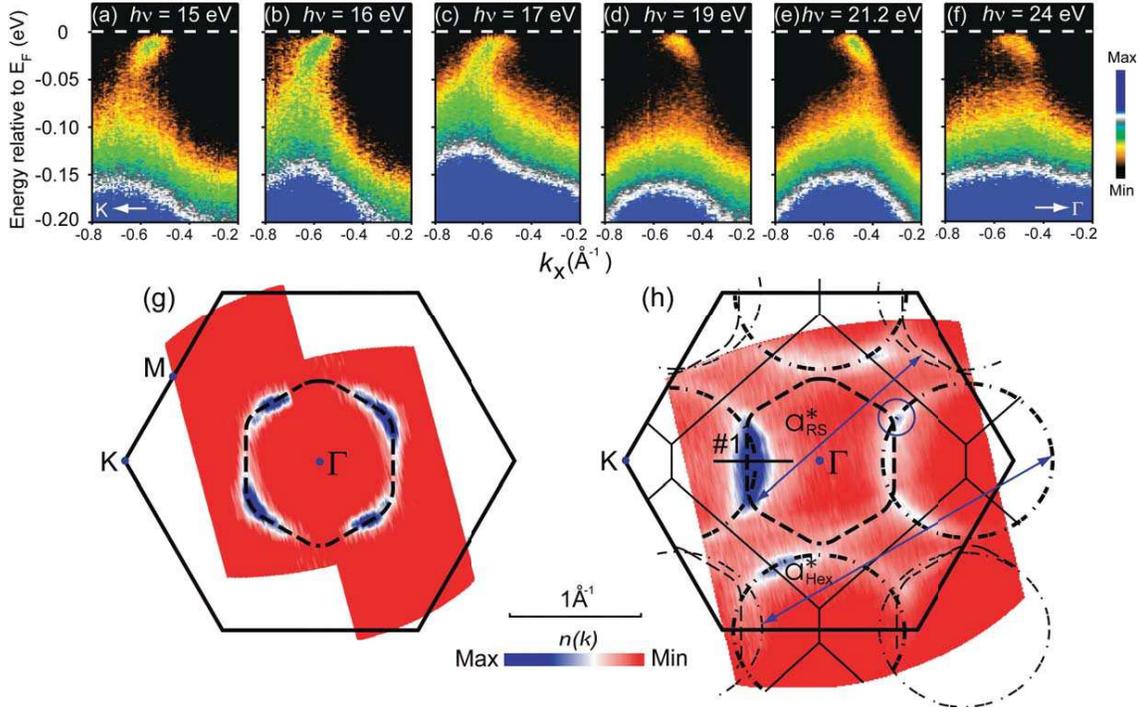}
\caption{(color online). ARPES data for
Bi$_{2}$Ba$_{1.3}$K$_{0.6}$Co$_{2.1}$O$_{y}$ along the cut \#1
direction with different photon energies: (a) \emph{h}$\nu$ =
15\,eV, (b) \emph{h}$\nu$ = 16\,eV, (c) \emph{h}$\nu$ = 17\,eV, (d)
\emph{h}$\nu$ = 19\,eV, (e) \emph{h}$\nu$ = 21.2\,eV, (f)
\emph{h}$\nu$ = 24\,eV.  (g-h) Fermi Surfaces measured at 20K with
15\,eV and 21.2\,eV  photons respectively, obtained by integration
of the spectral weight in a 20meV window around E$_{F}$. The
dash-dotted and dashed thick curves are [CoO$_2$]-derived and
[BiO/BaO]-derived Fermi pockets respectively. Their umklapp Fermi
surfaces are indicated by thinner curves.}
\end{figure*}

Angle resolved photoemission spectroscopy (ARPES) is one of the most
powerful techniques for studying electronic structure of solids and
thin films. Pioneering angle-integrated photoemission experiments
have been conducted on pulsed-laser deposited oxide
interfaces\cite{Fujimori1,Fujimori2,Fujimori3,Fujimori4}. However,
because the interface is buried far underneath the surface, and the
mean free path of photoelectrons is very short, photoemission
signals from the interface are very weak, which forbids high
resolution and momentum resolved measurements. The recent
synthesized misfit oxide single crystals provide an alternative
solution\cite{CaCoO,Misfit1}. Since each unit cell of a misfit oxide
is usually composed of two kinds of oxide layers with distinct space
group symmetries, it resembles a very clean artificial oxide
heterostructure and suitable for photoemission experiments. For
example, each formula unit of misfit cobaltites, which are known for
their anomalously large thermoelectric power, consists of several
rocksalt-type MO layers (M=Ca, Sr, Ba, Bi) and one CdI$_2$-type
hexagonal CoO$_2$ layer with edge-shared CoO$_6$ octahedra,
\textit{e.g.}, [BaO]$_2$[BiO]$_2$[CoO$_2$]$_{\alpha}$ as shown in
Fig.1(a). The two sublattices of rocksalt (RS) layers and hexagonal
layer possess incommensurate lattice constants along one Bi-O-Bi
axis, and a commensurate one along the other, causing large strain
and a global mismatch of the two lattices.

In this Letter, we report ARPES measurements of a misfit structured
oxide, Bi$_{2}$Ba$_{1.3}$K$_{0.6}$Co$_{2.1}$O$_{y}$ (BBKCO), which
reveals the detailed electronic structure of an oxide interface for
the first time. We find that large strain in the rocksalt layer
significantly raise its chemical potential and induces a large
electron transfer to the less strained CoO$_2$ layer. At the
presence of two incommensurate crystal fields, the low energy
electronic states of each individual layer are confined within
itself, preserving its symmetry; but they undergo umklapp scattering
by the incommensurate crystal field from the neighboring layer.
Furthermore, a  novel interfacial enhancement of electron-phonon
interactions (likely with interfacial phonons) is discovered. These
novel electronic properties observed in oxide interface depict a
detailed microscopic picture of various important processes that
could occur at oxide interfaces in general, and provide a foundation
for the future design of oxide-based devices.

Bi$_{2}$Ba$_{1.3}$K$_{0.6}$Co$_{2.1}$O$_{y}$ possesses the highest
thermoelectric power and high conductivity in its class, and thus
high figure of merit for thermoelectric applications. The single
crystals were prepared by the flux method described in detail
elsewhere\cite{Luo}. Its structure is the same as that in Fig.1(a),
with K$^{+}$ ions doped into the BaO layer. The two adjacent BiO
layers are weakly bound by the van der Waals force, providing a
stable BiO natural cleavage plane. This is shown by the rectangular
low energy electron diffraction (LEED) pattern in Fig.1(b). The
[BiO/BaO] layers are orthorhombic, the lattice constants along the
two Bi-O-Bi bond directions are $a_{RS}$= 5.031 ${\AA}$,
$b_{RS}$=5.683 ${\AA}$ respectively. $a_{RS}$ matches the distance
between the neighboring Co ions  along the same direction, while for
the perpendicular direction, $b_{RS}$=1.97 $b_{CoO_{2}}$, which is
collinear but aperiodic, causes the global misfit of the lattice.
The CoO$_2$ sublattice preserves the structure of its free standing
form, close to that in Na$_x$CoO$_2$\cite{NaCoO}. However, the
[BiO/BaO] sublattice is largely distorted compared with the BiO, BaO
layers in cuprate superconductors Bi$_2$Sr$_2$CaCu$_8$O$_{8+\delta}$
and YBa$_2$Cu$_3$O$_{7-x}$: it is squeezed by 6.4\% along the
$a_{RS}$ direction, while elongated by about 5.75\% along the
$b_{RS}$ direction. The Ba$^{2+}$ ions are further displaced from
the BaO plane towards the oxygen ions of the CoO$_2$ layer by
0.4$\AA$. These huge displacements and the misfit indicate large
strain in the rocksalt layer. The reduced Brillouin zones for the
individual [BiO/BaO] layers and CoO$_{2}$ layer are plotted in Fig.
1(c).


ARPES measurements were performed at beam line 9 of Hiroshima
Synchrotron Radiation Center,  beam line 5-4 of Stanford Synchrotron
Radiation Laboratory (SSRL), and beam line 21 of National
Synchrotron Radiation Research Center. The first two beamlines are
equipped with a Scienta R4000 analyzer, while the latter is equipped
with a SES200 analyzer. Typical energy and angular resolutions are
10\,meV and $0.3^\circ$ respectively in the experiments. The samples
were cleaved/measured in ultra-high vacuum ($\sim 5\times
10^{-11}\,mbar$).


Two kinds of states with distinct photon energy dependence are
identified for BBKCO. Fig.\,2(a-f) show the photoemission intensity
along the $\Gamma$-K direction taken with six different photon
energies between 15\,eV and 24\,eV. With relatively low energy
photons [Fig.2(a-c)], photoemission data show only one feature that
disperses towards K at higher binding energies. Correspondingly, the
map of photoelectron intensity at the Fermi energy (E$_{F}$) gives a
hexagonal hole-type Fermi pocket centered around $\Gamma$ in
Fig.2(g) (dashed lines). This resembles the Fermi surface observed
in Na$_x$CoO$_2$\cite{Yang}, whose six-fold symmetry clearly
demonstrates its CoO$_{2}$ origin. With higher energy photons,
another feature that disperses to the opposite direction becomes
dominant [Fig.2(d-f)]. Correspondingly in Fig.2(h), four hole-type
Fermi pockets show up in the intensity map measured at 21.2\,eV,
which could be fitted with the four dash-dotted ellipses. These
Fermi surfaces with the  symmetry of the rocksalt layer were not
observed in previous ARPES works of other misfit
cobaltites\cite{Brouet,Yusof}. Clearly, they should be mostly
originated from the [BiO/BaO] layers. As will be shown later, the
observation of the low energy electronic states in both layers
enables us to unveil various interfacial effects.


The van der Waals bond between two neighboring BiO layers suggests
weak inter-unit-cell coupling. Consistently, the Fermi crossing
momenta ($k_F$'s) in Fig.2(a-f) show negligible dependency on the
photon energies, which samples different out-of-plane momenta
$k_z$'s. For a two dimensional state, one can estimate its occupancy
through the Fermi surface volume based on the Luttinger theorem,
which robustly holds even in the presence of strong electron
correlation and/or electron-phonon interaction\cite{Dwshen}. The
CoO$_2$ layers are usually stoichiometric, which would be half
filled in undoped case. The actual Fermi surface volume indicates
that each CoO$_2$ formula unit has 0.6 extra electrons. Like in
Na$_x$CoO$_2$, the excess carriers come from outside, \textit{i.e.},
the [BiO/BaO] layers here. The [BiO/SrO] layers in cuprate
superconductors are prototypical charge reservoir, which donate
holes to the CuO$_2$ plane. Similarly, one would expect [BiO/BaO]
layers to donate holes, especially when its Ba$^{2+}$ ions are
replaced with  K$^+$ ions. However, in
Bi$_{2}$Ba$_{1.3}$K$_{0.6}$Co$_{2.1}$O$_{y}$, the Fermi surface
volume of the [BiO/BaO] rocksalt layers indicates that 0.6
\textit{electrons} are missing for each
[BiO/Ba$_{0.65}$K$_{0.3}$Co$_{0.05}$O] formula unit, which exactly
matches the additional electrons in the CoO$_2$ layer. This large
electron (instead of hole) loss out of the [BiO/BaO] layers clearly
demonstrates that, when huge strain is present, it would
significantly raise the energy of electronic states, and
consequently, the electrons flow into the low energy states of the
nearby layer. For BBKCO, this charge transfer would save the total
energy by hundreds of milli-electronvolt per chemical formula, and
thus greatly stabilize the misfit structure.\cite{estimation}

A charge transfer of $0.6\,e^{-}$ would induce a Madelung potential
as large as 24eV based on simple electrostatic
calculation\cite{OKAnderson}. Therefore, this unrealistic potential
must be screened by a ``back-flow'' of electrons of similar
magnitude at the valence bands\cite{ZYLu}; in other words, there is
covalent bonding across the interface. The evidence for such
covalent bonding is indeed found for high-lying states in their
dispersion along the out-of-plane dispersion. Fig.3 shows the
valence band dispersion taken along the $\Gamma$-K direction for
different $k_z$'s with four different photon energies. Six bands are
discernible in total, and the dispersions of bands below -2\,eV all
display some variations with $k_z$, with the largest dispersion of
about 0.5\,eV.  These bands are mainly of Ba, Bi and Oxygen
character based on a band structure calculation\cite{YZhu}. This
$k_z$ dispersion is quite remarkable, since it shows the long range
coherence along the c-axis is robust despite that the misfit would
cause the bond angle and length to vary along $c$ axis. In the end,
for a realistic Madelung potential difference of 1$\sim$2eV between
layers, the \textit{net} charge difference across the interface
would be less than 0.025$\sim$0.05 $e^{-}$.

\begin{figure}[t]
\includegraphics[width=8.5cm]{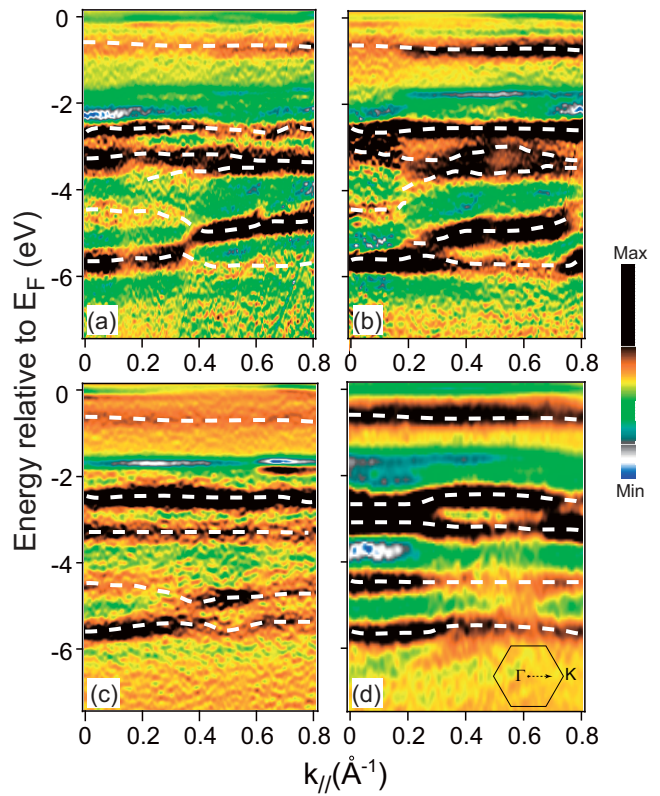}
\caption{ (color online). Valence band structure determined by
taking 2nd derivative of the photoemission intensity along the
$\Gamma$-K direction taken with (a) 17\,eV, (b) 21.2\,eV, (c)
28\,eV, and (d) 35\,eV photons respectively. The dashed lines are
identified band structure.}
\end{figure}


Large charge transfer might allude to strong couplings between two
layers, and consequently, bands in different layers would hybridize.
Particularly, signatures of band hybridization would show up in the
vicinity where their Fermi surfaces cross. On the contrary, we found
that in Fig.2(h) the CoO$_2$ Fermi surface crosses the rocksalt
layer Fermi surface without any sign of anti-crossing or bending. In
particular, there is even a slight enhancement of spectral weight at
the crossing momentum, as highlighted by the thin solid circle. The
observed two independent Fermi surface sets with distinct symmetries
prove that the low energy electronic states of [BiO/BaO] or CoO$_2$
layer are spatially confined within itself. This interesting finding
is further evidenced by the detailed properties of the
quasiparticles. Fig.4(a-b) show the photoemission images of the
quasiparticles in the CoO$_2$ layer, and the [BiO/BaO] layers
respectively along the $\Gamma-K$ direction.  As shown in Fig.4(d),
the measured quasiparticle scattering rate is a linear function of
the binding energy for CoO$_2$ state, while it is a quadratic
function near $E_F$ for [BiO/BaO] state. The absence of
hybridization between states in two metallic layers of such an oxide
interface is quite unexpected, and very different from the strong
hybridization observed for quantum well states at
metal/semiconductor interfaces\cite{Chiang}.


Very interestingly, although the coupling between the low energy
electronic states are weak, interlayer interactions could still
manifest themselves in other forms,  for example, in the several
weak Fermi surfaces (thin lines) in Fig.2(h). If one would displace
the main Fermi surface of [BiO/BaO] layers by the reciprocal lattice
constant $a^{*}_{Hex}$ of the CoO$_2$ layer (illustrated by the
double-headed arrows), the resulting umklapp Fermi surfaces
perfectly fit the weak features in the data. Similarly, the CoO$_2$
umklapp Fermi surfaces are observed apart from the main ones by the
reciprocal lattice constant $a^{*}_{RS}$ of the [BiO/BaO] layers.
Therefore, this observation proves that the crystal field from one
side of the interface is imposed on the other, and act as an
incommensurate potential that scatters the electrons there.

\begin{figure}[t!]
\includegraphics[width=8.5cm]{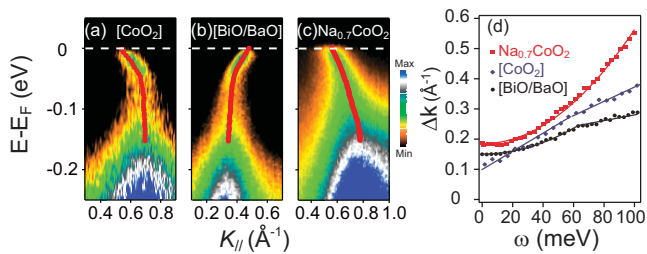}
\caption{(color online). Photoemission intensity along $\Gamma-K$
taken at 20K, for (a) BBKCO's  CoO$_2$ layer (\emph{h}$\nu$=15eV),
(b) BBKCO's [BiO/BaO] layers (\emph{h}$\nu$=21.2eV), and (c)
Na$_{0.7}$CoO$_2$ (\emph{h}$\nu$=21.2eV). The thick curves are
fitted dispersions. (d) The width of the momentum distribution curve
vs. binding energy for data presented in panel a-c, which is an
estimate of the quasiparticle scattering rate.}
\end{figure}

More interfacial effects could be revealed by comparing the BBKCO
data with the quasiparticle behavior of Na$_{0.7}$CoO$_2$ taken in
the same momentum region. While the Na$_{0.7}$CoO$_2$ dispersion is
a smooth curve [Fig.4(c)], there are strong kinks on the dispersions
of the BBKCO bands [Fig.4(a-b)]. These kinks are signature of
interaction with some bosonic modes, most likely phonons here,
considering the non-magnetic nature of the rocksalt layer. The
similar energy scale of 60\,meV for both CoO$_2$ and [BiO/BaO]
states indicates they might interact with the same phonons induced
at the highly strained interface. This strong coupling to a mode is
also reflected in its scattering rate [Fig.4(d)], where a clear
turning point around 60meV appears for both the bands of BBKCO.
Furthermore, the scattering rate of Na$_{0.7}$CoO$_2$ is a simple
quadratic function of binding energy, different from the linear
behavior of the BBKCO CoO$_2$ layer. This suggests that the
electron-phonon interactions induced by the interface may change the
quasiparticle behavior.


The current findings for an oxide interface provide some important
guidelines for future device design. For example, our results
suggest that interfacial strain might cause large charge transfer
without severely changing the  properties of participating layers,
since the coupling between the low energy electronic states could
still be weak. Meanwhile, the interfacial umklapp scattering and
phonon effects might alter the low energy electronic behavior.

To summarize, we have revealed various interfacial electronic
properties that have never been observed in oxide interfaces before.
We show that while the low energy states are confined within
individual layer, the high lying states are covalently coupled. The
relation between strain and charge transfer across oxide interface
is illustrated from a microscopic level. The enhancement of electron
phonon coupling and interfacial umklapp scattering are discovered.
Our findings provide an electronic structure foundation for
understanding oxide interfaces and designing oxide devices.

We gratefully acknowledge the experimental help from Dr. D. H. Lu,
and R. H. He during the SSRL experiment, and the helpful discussions
with Profs. Z.-X. Shen, C. Kim, Z. Q. Wang and Dr. T. Cuk. This work
was supported by NSFC, MOST (973 projects No.2006CB921300 and
2006CB922005), and STCSM of China. Portions of this research were
carried out at SSRL, a national user facility operated by Stanford
University on behalf of the U.S. DOE, Office of BES.


\begin{references}
\bibitem{estimation} The energy saved through charge transfer can be roughly estimated to be $v_{F}^{0} \times \Delta k$,
where $v_{F}^{0}\approx 1.4eV\AA$ is the bare band velocity of the
rocksalt layer, and $\Delta k\approx 0.35\AA^{-1}$ is the difference
of [BiO/BaO] Fermi pocket radius before and after the charge
transfer.
\end{references}
\end{document}